\documentclass[aps,prd,twocolumn,showpacs,superscriptaddress,groupedaddress,10pt]{revtex4-2}
\usepackage{graphicx}  
\usepackage{dcolumn}   
\usepackage{bm}        
\usepackage{amssymb}   

\usepackage{amsfonts}
\usepackage{amsmath}
\usepackage{bm}
\usepackage{array}
\usepackage{amssymb}
\usepackage{array}
\usepackage{amsthm}
\usepackage{amssymb}
\usepackage{braket}
\usepackage{color}

\newcommand{\be}{\begin{eqnarray}}
\newcommand{\ee}{\end{eqnarray}}
\renewcommand{\d}{\mbox{${\rm d}$}} 

\newcommand{\K}{\mathcal{K}}
\newcommand{\T}{\mathcal{T}}
\DeclareMathOperator{\arcsinh}{arcsinh}
\newtheorem{theorem}{Theorem}

\newtheorem{prop}{Proposition}
\newtheorem{lemma}{Lemma}

\theoremstyle{definition}

\theoremstyle{remark}

\numberwithin{equation}{section}
\begin{document}

\title{Thermal attractor to General Relativity in {\em vacuum} Induced Gravity}
\author{Andrea Giusti}
\affiliation{DIFA \& ${\cal AM}^2$, Università di Bologna, 40126 Bologna, Italy}
\affiliation{I.N.F.N., Sezione di Bologna, I.S.~FLAG}
\author{Gaetano Fiore}
\affiliation{Dipartimento di Matematica ``Renato Caccioppoli'', Università di Napoli ``Federico II'', Naples, Italy}
\affiliation{I.N.F.N., Sezione di Napoli}
%
%

\begin{abstract}
\noindent We investigate the attractor-to-General-Relativity mechanism in {\em vacuum} induced gravity within spatially-flat, expanding, homogeneous, and isotropic cosmology. Taking advantage of the equivalence with Brans--Dicke gravity with positive constant coupling and a quadratic potential, we provide a proof of future global existence and convergence to a family of de Sitter solutions. Then, for a future-directed scalar-field gradient, we show that the evolution remains in the ``cooling region'' of the thermodynamic expansion-temperature diagram and approaches the zero-temperature GR equilibrium. These results corroborate the thermodynamic description of the attractor to GR in this setting and complement earlier investigations based on dynamical systems and exact solutions.
\end{abstract}

\maketitle

\section{Introduction}
\label{sec:intro}
Induced gravity \cite{Sakharov:1967nyk, Zee:1978wi, Cooper:1981byv, Zee:1980sj, Turchetti:1981zb} consists in a ``first-generation'' globally scale-invariant
scalar-tensor theory defined by the action 
\be
\label{eq:CV}
\mathcal{S}[g_{ab},\sigma, \zeta] &\!=\!& \int \d ^4 x \sqrt{-g} 
\left[ \frac{\xi}{2} \, \sigma^2 \, R - \frac{1}{2} \, \nabla_a \sigma \nabla^a \sigma - \frac{\lambda}{4} \, \sigma^4 \right] \nonumber \\
&& + \mathcal{S}^\mathrm{(m)} [g_{ab},\zeta] \, , \nonumber \\
&& 
\ee 
with $\lambda, \xi > 0$ being constant and dimensionless, $\sigma$ denoting a scalar field, $g$ representing the determinant of the spacetime metric $g_{ab}$ with Ricci scalar $R$, $\nabla$ being the Levi-Civita connection, and $\mathcal{S}^\mathrm{(m)}$ accounting for the contribution of matter fields  (collectively denoted by) $\zeta$. Note also that in this work we adopt the notation of Ref.~\cite{Wald:1984rg}, in which the metric signature is $(-+++)$. Furthermore, in our units we have that the speed of light and $8 \pi G$, where $G$ denotes Newton's constant, are unity.

This theory is known to map onto the following realization of Brans--Dicke (BD) gravity (see e.g. \cite{Finelli:2007wb})
\be 
\label{eq:BDA}
\mathcal{S}[g_{ab},\phi, \zeta] &\!=\!& \frac{1}{2}\int \d ^4 x \sqrt{-g} 
\left[ \phi \, R - \frac{\nabla_a \phi \nabla^a \phi}{4\xi \,\phi} - \frac{\lambda \, \phi^2}{2 \xi^2} \right] \nonumber \\ 
&& + \mathcal{S}^\mathrm{(m)} [g_{ab},\zeta]\, ,
\ee
{\em via} the simple field redefinition $\phi = \xi \sigma^2$. In other words, induced gravity corresponds to BD gravity with BD potential and coupling that read
\be 
\label{eq:PotAndOmega}
V(\phi) \equiv V_0 \, \phi^2 = \frac{\lambda}{2 \xi^2} \, \phi^2 \quad \mbox{and} \quad 
\omega = \frac{1}{4 \xi} \, .
\ee
As pointed out in \cite{Giusti:2026ymb}, the positivity of the induced gravity constants allows the theory to avoid pathologies such as phantom phenomenology
and the loss of the dynamics of the BD scalar $\phi$.

In this work we focus on {\em vacuum} cosmological solutions of induced gravity (i.e. $\mathcal{S}^\mathrm{(m)} [g_{ab},\zeta] = 0$). Proceeding in analogy with \cite{Giusti:2026ymb}, we consider a flat Friedmann-Lema\^{i}tre-Robertson-Walker (FLRW) spacetime, with line element
$$
\d s^2 = - \d t^2 + a^2(t) \left( \d r^2 + r^2 \d \Omega^2 \right)
$$
where $t$ is cosmic time, $a(t)$ is the scale factor, $r$ denotes the comoving radial coordinate, and $\d \Omega^2$ is the line element on the unit $2$-sphere. Then, the evolution equations are given by 
the {\em BD scalar field equation}
\be 
\ddot{\phi} + 3H \dot{\phi} = 0 \, , \label{eq:scalar} 
\ee
where $\phi = \phi (t)$ due to homogeneity and isotropy, and the {\em modified Friedmann equation}
\be 
H^2 = \frac{\omega}{6} \frac{\dot{\phi}^2}{\phi^2} - H \frac{\dot{\phi}}{\phi} + \frac{V_0 \, \phi}{6} \label{constraint} \, ,
\ee
with $H := \dot{a}/a$, the Hubble parameter, and where the dot denotes the derivative with respect to $t$. Note that the Raychaudhuri (acceleration) equation is redundant.

It is now trivial to show, as observed in \cite{Cooper:1981byv,Faraoni:2004pi}, 
that the system consisting of Eqs.~\eqref{eq:scalar} and \eqref{constraint}
does not admit any power-law solutions except for the trivial ones with
\be 
\label{eq:dSstate}
\phi = \phi_{\rm c} \quad \mbox{and} \quad H^2 = \frac{\lambda \, \phi_{\rm c}}{12 \, \xi^2} \, ,
\ee 
for any $\phi_{\rm c} > 0$ constant, which is nothing but a class of de Sitter (dS) solutions parametrised by $(\phi_{\rm c},\xi, \lambda)$ (see also \cite{Cooper:1981byv,Turchetti:1981zb}). The fact that $\phi = \phi_{\rm c} = {\rm constant}$ tells us that these dS solutions lie at the intersection of BD gravity and General Relativity (GR). Specifically, for these solutions BD gravity is reduced to GR plus a cosmological constant $\Lambda = 3 \, H^2 = \lambda \, \phi_{\rm c}/(4 \, \xi^2)$.

The paper is organized as follows. In Section~\ref{sec:ThermoIG} we review the thermodynamic description of scalar-tensor gravity and its specialization to induced gravity. In Section~\ref{sec:dS} we establish a future global-existence result and convergence to a de Sitter state, and obtain quantitative estimates for the approach to equilibrium. In Section~\ref{sec:discussion} we discuss these results within the formalism of the thermodynamics of scalar-tensor gravity and discuss the attractor-to-GR mechanism in the $(\Theta,\mathcal{K}\mathcal{T})$--plane. Finally, Section~\ref{sec:conclusions} we present some concluding remarks.

\section{Thermal view of Induced Gravity}
\label{sec:ThermoIG}
Here we briefly review the generalities of the thermodynamics 
of scalar-tensor theories, with particular regard for its specialization to induced gravity. For further details on this formalism we refer the reader to
\cite{Faraoni:2018qdr,Faraoni:2021lfc,Faraoni:2021jri,Giusti:2021sku,Giardino:2023ygc,Banerjee:2026ulx,Miranda:2026ozq}. We will be particularly interested in the case of cosmology, hence in this context we refer the reader to \cite{Giardino:2022sdv, Miranda:2024dhw,Faraoni:2025alq} for further details. Notably, Ref.~\cite{Faraoni:2025alq} represents the main motivation for the present investigation and concerns a thermal description of the attractor-to-GR mechanism in scalar-tensor gravity, originally proposed by T.~Damour and
K.~Nordvedt in  \cite{Damour:1992kf,Damour:1993id}.

Since the induced gravity model can be mapped onto a specific realization of BD gravity, we can discuss its thermal view, first formalized in detail in \cite{Giusti:2026ymb}, by restricting it to this well-established setting.

As discussed in \cite{Faraoni:2018qdr} (see also \cite{Deffayet:2010qz,Pujolas:2011he}), certain scalar-tensor theories admit an imperfect fluid representation. Specifically, the effective Einstein equation of BD gravity can be written implicitly as \cite{Faraoni:2018qdr}
\begin{equation}
\label{eff-efe}
    G_{ab} = G_{\rm eff} (\phi) \, T^{\rm (m)} _{ab} + T^{\rm (eff)} _{ab} \, ,
\end{equation}
where $G_{ab}$ is the Einstein tensor, $\phi > 0$ denotes the BD scalar field, $G_{\rm eff} (\phi) = 1/\phi$ is an effective gravitational coupling, and $T^{\rm (eff)} _{ab}$ is a rank-2 symmetric tensor field containing information about the remaining contributions of the BD field (and its derivatives) to the effective Einstein equation.

The underlying idea behind the imperfect fluid approach to modified gravity is to treat $T^{\rm (eff)} _{ab}$ as a stress-energy tensor of an effective fluid. This can be done in a natural way if $\nabla^a \phi$ is timelike. Indeed, in this case the vector field
\begin{equation}
\label{eq:ua}
u^a := \epsilon \, \frac{\nabla^a\phi}{\sqrt{2X}} \, , \quad 2X := -\nabla^e \phi \nabla_e\phi \, ,
\end{equation}
where $\epsilon = \pm 1$ has to be chosen so that $u^a$ is future-oriented by definition, can be interpreted as the velocity of an observer comoving with the effective fluid \cite{Faraoni:2018qdr}. 

Let us assume $\nabla^a \phi$ is timelike, which is indeed the case of interest for cosmology. Once $u^a$ is defined as in \eqref{eq:ua}, the effective fluid described by the stress-energy tensor $T^{\rm (eff)} _{ab}$ admits (see \cite{Faraoni:2023hwu}) an imperfect fluid decomposition \cite{Faraoni:2018qdr,Pujolas:2011he} 
\be
T^{(\rm eff)} _{ab} =\rho \, u_a u_b 
+P \, h_{ab} +\pi_{ab} + 2 \, q_{(a} u_{b)} 
\,, 
\label{eq:imperfect}
\ee
with $\rho$, $P$, $\pi_{ab}$ and $q_a$ denoting, respectively, the effective energy density,  isotropic pressure, anisotropic stress tensor, heat flux density, $h_{ab} := g_{ab}+u_au_b$ is the induced metric onto the $3$-space orthogonal to $u^a$, and indices enclosed in parentheses are understood to be symmetrized. 

Note that the explicit functional dependence of $\rho$, $P$, $\pi_{ab}$ and $q_a$ on the scalar field and its derivatives have been derived in \cite{Faraoni:2018qdr,Faraoni:2021lfc,Faraoni:2021jri}, and hence we refer the reader to these works for further details.

It was then shown \cite{Faraoni:2021lfc,Faraoni:2021jri} that comparing the expressions of the effective imperfect-fluid components of  $T^{(\rm eff)} _{ab}$ discussed above with the kinematic quantities associated with the fluid lines (computed from the covariant velocity gradient) allows one to define a map between the constitutive laws of the effective fluid and a specific realization Eckart's theory \cite{Eckart:1940te}. This was done for the specific case of BD gravity in \cite{Faraoni:2021lfc,Faraoni:2021jri}, and was later extended to viable Horndeski gravity in \cite{Giusti:2021sku,Miranda:2024dhw}.

This formal analogy with Eckart's theory yields a notion of temperature $\T$ (multiplied by the thermal conductivity of the fluid $\K$) 
of modified gravity that, in the case of BD theory, reads \cite{Faraoni:2021lfc,Faraoni:2021jri}
\be \label{KTdefinition}
\K \T = \epsilon \, \frac{\sqrt{2X}}{\phi} \, .
\ee
Since GR is obtained from BD theory by setting $\phi = {\rm constant}$, it corresponds to a zero-temperature equilibrium state in this thermodynamic framework. Hence, in general the value of $\K\T$ measures the departure of a given scalar–tensor configuration from the GR equilibrium (see \cite{Faraoni:2021lfc,Faraoni:2021jri,Giusti:2021sku,Giardino:2023ygc,Giusti:2021sku,Miranda:2024dhw}). It is also worth pointing out that \eqref{KTdefinition} is not the only possible definition of temperature compatible with the analogy with Eckart's theory \cite{Gallerani:2024gdy}; however, the resulting alternative formalism is physically equivalent to the one presented here.

An important consequence of this formalism is that it allows one to determine an evolution equation for the temperature, often referred to as the ``generalised heat equation'' \cite{Faraoni:2021jri,Faraoni:2021lfc}. Specifically, in the case of BD gravity this equation reads \cite{Faraoni:2021lfc,Faraoni:2021jri}
\begin{equation}
\label{eq:heat}
\frac{\d\left( \K\T \right)}{\d\tau} := u^c \nabla_c \left( \K\T \right) = \K\T \left( \K\T -\Theta \right) + \frac{\Box \phi}{\phi} \, ,
\end{equation}
where $\Box \phi := \nabla_c \nabla^c \phi$, $\Theta := \nabla_a u^a$ denotes the expansion scalar, and $\tau$ is the proper time along the fluid lines.

Starting from \eqref{eq:heat} it was possible to provide a thermodynamic derivation of the attractor-to-GR mechanism for ``first-generation'' scalar-tensor theories (see \cite{Faraoni:2025alq}), which include BD theory as a subclass. As a consequence, restricting our attention to BD gravity with conformal matter ($T^{(m)}=0$), constant coupling $\omega$, and a quadratic potential, Eq.~\eqref{eq:heat} reduces to
\begin{equation} \label{eq:heat-vacuum}
\frac{\d(\K\T)}{\d\tau}
=\K\T(\K\T-\Theta) \, ,
\end{equation}
so an expanding solution evolves away from GR, in the sense of increasing $\K\T$, as long as $\K\T>\Theta>0$. This analysis can be easily extended to the remaining three quadrants of the $(\Theta, \K\T)$--plane.

\section{Late-time de Sitter}
\label{sec:dS}
Dynamical system methods for studying attractor solutions are commonly employed in scalar-tensor cosmology (see e.g. \cite{Faraoni:2004pi,Bahamonde:2017ize} and references therein). In this section we provide a simple (pedagogical) proof of the global existence of expanding solutions of the system \eqref{eq:scalar}--\eqref{constraint}, and its convergence toward a de~Sitter configuration of the type in \eqref{eq:dSstate} at late (cosmic) times, by taking advantage of standard methods from the theory of ordinary differential equations (see specifically \cite{Teschl2012}). Note that some precise investigations of the de Sitter (dS) attractor in induced gravity (or its equivalent BD version) have appeared in the literature \cite{Faraoni:2004dn, Faraoni:2006ik, Cervantes-Cota:2010aau,Faraoni:2012bf,Hrycyna:2013yia,Hrycyna:2013hla}.

We begin by considering the modified Friedmann equation \eqref{constraint} as a algebraic second-degree equation in $H$. This implies the following
\begin{lemma}
Let $\phi > 0$, then
\begin{equation}
    \label{eq:H}
    H := - \frac{\dot{\phi}}{2\phi} + \frac{1}{2} \sqrt{\left( 1 + \frac{2}{3} \omega \right) \left(\frac{\dot{\phi}}{\phi} \right)^2 + \frac{2}{3} \, V_0 \, \phi} \, ,
\end{equation}
is a positive definite solution of \eqref{constraint}.
\end{lemma}
\begin{proof}
Treating Eq.~\eqref{constraint} as a second degree equation in $H$, we have that the discriminant reads:
$$
\Delta = \left( 1 + \frac{2}{3} \omega \right) \left(\frac{\dot{\phi}}{\phi} \right)^2 + \frac{2}{3} \, V_0 \, \phi 
\, ,
$$
which is always strictly positive due to our assumptions on $\phi$ and the parameters $V_0$ and $\omega$, moreover since $(2/3) \, V_0 \, \phi > 0$ we also have that
$$
\Delta > \left( 1 + \frac{2}{3} \omega \right) \left(\frac{\dot{\phi}}{\phi} \right)^2  \, .
$$
Then, solving Eq.~\eqref{constraint} for $H$ yields
$$
 H_{\pm} = - \frac{\dot{\phi}}{2\phi} \pm \frac{1}{2} \sqrt{\Delta (\phi, \dot{\phi})} \, ,
$$
which shows that \eqref{eq:H} indeed represents one branch, i.e. $H_+$, of the solution of Eq.~\eqref{constraint}. Moreover, from the previous inequality for $\Delta$ we have that
$$
\sqrt{\Delta} > \left|\dot{\phi}/\phi\right| \, ,
$$
and hence that $H_{+} > 0$, $\forall \dot{\phi} \in \mathbb{R}$, which concludes the proof.
\end{proof}

Selecting this branch \eqref{eq:H} we are therefore restricting our analysis to the case of an {\em expanding universe}. The next step is to characterize the evolution of the scalar field $\phi (t)$ given Eq.~\eqref{eq:H} and the initial conditions of interest for the thermal analogy.

We begin with two preparatory results.
\begin{lemma} \label{vneg}
Consider the scalar field equation \eqref{eq:scalar} with initial conditions $\phi (0) = \phi_0 > 0$, $\dot\phi (0) = v_0 \in \mathbb{R}$, and $a(0) = 1$. Then one has that the time derivative of the scalar field reads
\begin{equation}
\label{eq:dotphi}
\dot{\phi}(t) = \frac{v_0}{a^3 (t)} \, , 
\end{equation}
$\forall t \geq 0$ for which the solution of the system \eqref{eq:scalar}--\eqref{constraint} exists.
\end{lemma}

\begin{proof}
We can solve Eq.~\eqref{eq:scalar}, with respect to $\dot{\phi}$, by separation of variables, i.e.
$$
\ddot{\phi} + 3H \, \dot{\phi} = 0 \,\,\, \Rightarrow \,\,\, 
\frac{\d}{\d t}\!\left(a^3\dot\phi\right)=0
\,\,\, \Rightarrow \,\,\, \dot{\phi} = \frac{C}{a^3(t)} \, ,
$$
with $C$ an integration constant. If we now impose the initial conditions $\dot{\phi}(0) = v_0 \in \mathbb{R}$ and $a(0) = 1$, the result follows immediately.
\end{proof}

Let us rewrite \eqref{eq:scalar}--\eqref{constraint} as the first-order system:
\begin{equation}
\label{eq:System}
\left\{
\begin{aligned}
& \dot{\phi} = v \, ,\\
& \dot{v} = - 3 H(\phi,v) \, v \, ,
\end{aligned}
\right.
\end{equation}
with $H$ as in \eqref{eq:H} and initial conditions 
\begin{equation}
\label{eq:ICs}
\left\{
\begin{aligned}
& \phi(0) = \phi_0 > 0 \, , \\
& v (0) = v_0 \in \mathbb{R} \, .
\end{aligned}
\right.
\end{equation}
It is clear that this system of ordinary differential equations is singular for $\phi = 0$, since $H(\phi,v)$ is not defined for $\phi = 0$. 

We can now immediately spot that the defining vector field of the first-order system \eqref{eq:System}, i.e.
$$
F ( \phi , v) := 
\begin{pmatrix}
v \\
- 3Hv 
\end{pmatrix} \, ,
$$
is $C^\infty (\mathcal{D},\mathbb{R}^2)$ with $\mathcal{D} := (0,+\infty)\times\mathbb R$, and hence $F$ is locally Lipschitz on $\mathcal{D}$. The Picard--Lindel{\"o}f theorem \cite{Teschl2012} ensures the local existence and uniqueness for every initial datum in $D$. By the maximal extension theorem \cite{Teschl2012} we hence have that for each initial datum in $\mathcal{D}$, there exists a unique maximal solution of \eqref{eq:System}, whose maximal future interval of existence is $[0,T_\ast)$, with $T_\ast\in(0,+\infty]$.

Interestingly, we can manipulate the autonomous system \eqref{eq:System} to derive a {\em first integral} \cite{Teschl2012}, i.e. a nontrivial continuously differentiable function that remains constant along every solution.  

By the inverse function theorem, if $\dot{\phi} \neq 0$ then $t\mapsto\phi(t)$ is locally invertible and, by the chain rule, we get
\begin{equation}
\frac{\d\dot{\phi}}{\d\phi}
=\frac{\d}{\d\phi}\left[\dot\phi(t(\phi))\right]
=\frac{\ddot\phi}{\dot\phi} \, .
\end{equation}
From \eqref{eq:scalar} and \eqref{eq:H} we therefore find that
\begin{eqnarray}
    \frac{\d\dot{\phi}}{\d\phi} &=& \frac{\ddot\phi}{\dot\phi} = - 3H\\
    &=& \frac{3 \,\dot{\phi}}{2 \, \phi} - \frac{3}{2} \sqrt{\left( 1 + \frac{2}{3} \omega \right) \left(\frac{\dot{\phi}}{\phi} \right)^2 + \frac{2}{3} \, V_0 \, \phi} \, , \qquad
\end{eqnarray}
which can be rewritten as
\begin{eqnarray}
    \label{eq:dotphionphi}
  \frac{\d\dot{\phi}}{\d\phi} - \frac{3 \,\dot{\phi}}{2 \, \phi} &=& 
  -\frac{3}{2} \sqrt{\left( 1 + \frac{2}{3} \omega \right) \left(\frac{\dot{\phi}}{\phi} \right)^2 + \frac{2}{3} \, V_0 \, \phi} \, . \nonumber \\ 
  & &
\end{eqnarray}
A straightforward computation shows that the left-hand side of Eq.~\eqref{eq:dotphionphi} can be rewritten as
\begin{eqnarray}
\frac{\d\dot{\phi}}{\d\phi} - \frac{3 \,\dot{\phi}}{2 \, \phi} = \phi^{3/2} \, \frac{\d}{\d \phi} \left( \frac{\dot{\phi}}{\phi^{3/2}} \right) \, ,   
\end{eqnarray}
and hence \eqref{eq:dotphionphi} becomes
\begin{eqnarray}
    \phi^{3/2} \, \frac{\d}{\d \phi} \left( \frac{\dot{\phi}}{\phi^{3/2}} \right) &=& -\frac{3}{2} \sqrt{\left( 1 + \frac{2}{3} \omega \right) \left(\frac{\dot{\phi}}{\phi} \right)^2 + \frac{2}{3} \, V_0 \, \phi} \, , \nonumber \\
    &&
\end{eqnarray}
that, following simple manipulations, implies
\begin{eqnarray}
    \label{eq:int-1}
    \frac{\d}{\d \phi} \left( \frac{\dot{\phi}}{\phi^{3/2}} \right) &=& 
    -\frac{3}{2 \, \phi} \sqrt{\left( 1 + \frac{2}{3} \omega \right) \left(\frac{\dot{\phi}}{\phi^{3/2}} \right)^2 + \frac{2}{3} \, V_0} \, . \nonumber \\
    &&
\end{eqnarray}
Introducing the auxiliary variable $X := \dot{\phi} / \phi^{3/2}$, Eq.~\eqref{eq:int-1} becomes
\begin{eqnarray}
    \label{eq:int-2}
    \frac{\d X}{\d \phi} = - 
    \frac{3}{2 \, \phi} \sqrt{\left( 1 + \frac{2}{3} \omega \right) X^2 + \frac{2}{3} \, V_0} \, ,
\end{eqnarray}
which can be easily integrated by separation of variables. Indeed, we have that \eqref{eq:int-2} can be rewritten as
\begin{eqnarray}
 \frac{\d X}{\sqrt{\left( 1 + \frac{2}{3} \omega \right) X^2 + \frac{2}{3} \, V_0}} = - 
    \frac{3}{2} \frac{\d \phi}{\phi} \, ,
\end{eqnarray}
and hence
\begin{eqnarray*}
\frac{\arcsinh\!\left(
\sqrt{\frac{3+2\omega}{2V_0}}\,X
\right)}{\sqrt{1+\frac{2}{3}\omega}} = - \frac{3}{2} \log \left( \frac{\phi}{\phi_0}\right) + C \, ,
\end{eqnarray*}
with $C$ a constant, $\phi_0 = \phi (0)$, and $\arcsinh$ denoting the inverse hyperbolic sine. Hence, we have that the function
\begin{eqnarray} \label{eq:first-int}
I(\phi,v) &=& \frac{\arcsinh\!\left(
\sqrt{\frac{3+2\omega}{2V_0}}\,\frac{v}{\phi^{3/2}}
\right)}{\sqrt{1+\frac{2}{3}\omega}} + \frac{3}{2} \log \left( \frac{\phi}{\phi_0}\right) \, , \nonumber \\
& & 
\end{eqnarray}
is both constant along every solution trajectory of \eqref{eq:System} and continuously differentiable on $\mathcal D$. Thus $I(\phi,v)$ is a first integral for the system \eqref{eq:System}.

Since $I(\phi,v)$ is constant along the orbits of the system, its value will be set by the initial conditions \eqref{eq:ICs}. Indeed, along the orbits of the system \eqref{eq:System} we have that 
$$
\dot{I} (t) \equiv \frac{\d}{\d t} I (\phi (t),v(t)) = 0 \, ,
$$
and hence 
$$
I (\phi (t),v(t)) = I (\phi_0,v_0) = 
\frac{\arcsinh\!\left(
\sqrt{\frac{3+2\omega}{2V_0}}\,\frac{v_0}{\phi_0^{3/2}}
\right)}{\sqrt{1+\frac{2}{3}\omega}} \, .
$$

If we now introduce the quantity
\begin{equation}
\begin{split}
    \phi_\infty &:= 
    \phi_0 \,{\rm e}^{\frac{2}{3}
    I (\phi_0,v_0)}\\
    &= 
    \phi_0 \,\exp\!\left[\frac{2}{3}
\frac{\arcsinh\!\left(
\sqrt{\frac{3+2\omega}{2V_0}}\,\frac{v_0}{\phi_0^{3/2}}
\right)}{\sqrt{1+\frac{2}{3}\omega}}\right] \, ,
\end{split}
\end{equation}
which is a strictly positive quantity, we find the following result. 

\begin{prop} \label{prop-bounds}
    For the maximal solution of the system \eqref{eq:System} the following conditions hold:
    \begin{equation}
        \begin{array}{lll}
{\rm If} \,\,v_0 < 0:&\phi_\infty<\phi(t)\leq\phi_0,&v_0\leq\dot{\phi}(t)<0,\\[2pt]
{\rm If} \,\, v_0=0:&\phi(t)=\phi_\infty=\phi_0,&\dot{\phi}(t)=0,\\[2pt]
{\rm If} \,\, v_0>0:&\phi_0\leq\phi(t)<\phi_\infty,&0<\dot{\phi}(t)\leq v_0 \, ,
\end{array}
\label{eq:compactness}
    \end{equation}
for all $t \in [0,T_\ast)$.
\end{prop}
\begin{proof}
    Let us introduce the notation
    $$
    A(t) \equiv A(\phi (t),v(t)) := \frac{\arcsinh\!\left(\sqrt{\frac{3+2\omega}{2V_0}}\,\frac{v(t)}{\phi(t)^{3/2}} \right)}{\sqrt{1+\frac{2}{3}\omega}} \, . 
    $$
    Consider the condition $I (\phi (t),v(t)) = I (\phi_0,v_0)$. According to the notation above then the condition can be rewritten as 
    $$
    \frac{3}{2} \log \left( \frac{\phi}{\phi_0}\right) = A(0) - A(t) \, .
    $$
    
    If $v_0 < 0$, by Lemma~\ref{vneg} we know that $v = \dot{\phi} < 0$ and $v_0\leq\dot{\phi}(t)<0$ for an expanding universe. Using the fact that $\arcsinh$ is an odd function, we have that $A(t) < 0$. This implies that
    $$
    \frac{3}{2} \log \left( \frac{\phi}{\phi_0}\right) = A(0) - A(t) > A(0) \, ,
    $$
    and hence 
    $$
    \phi > \phi_0 \, {\rm e}^{\frac{2}{3} A(0)} = \phi_0 \, {\rm e}^{\frac{2}{3} I(\phi_0,v_0)} = \phi_\infty \, .
    $$
    Furthermore, Lemma~\ref{vneg} also implies that, for $v_0 < 0$, $\phi$ is strictly monotonically decreasing, and hence 
    $$\phi(t) \leq \phi(0) = \phi_0 \, , \qquad \mbox{for } \,  t \in [0,T_\ast) \, ,$$
    which concludes the proof of the first point.

    If $v_0 > 0$, the same argument above applies with the inequality reversed. Lastly, the case $v_0 = 0$ is trivial.
\end{proof}

In other words, we have found that $\phi_\infty$ acts as a barrier that $\phi(t)$ cannot cross during the evolution governed by \eqref{eq:System}.

We can now prove the a (future) global existence result for the solution of \eqref{eq:System} under the conditions \eqref{eq:ICs}, together with the asymptotic convergence to the GR--dS limit \eqref{eq:dSstate} of induced gravity.

\begin{theorem} \label{th-sol}
    Let $\omega, V_0, \phi_0 > 0$ and $v_0 \in \mathbb{R}$. The first-order system \eqref{eq:System}, with initial conditions \eqref{eq:ICs}, has a unique solution $\forall t \geq 0$. Furthermore, this solution is such that
    \begin{eqnarray}
        \lim_{t \to + \infty} \phi (t) &=& \phi_\infty \, , \label{lim-phi}\\
        \lim_{t \to + \infty} \dot{\phi} (t) &=& 0 \, , \label{lim-dphi}\\
        \lim_{t \to + \infty} H (t) &=&  \sqrt{\frac{V_0 \, \phi_\infty}{6}} =:H_\infty \, . \label{lim-H}
    \end{eqnarray}
\end{theorem}
\begin{proof}
Let $m_\phi := \min\{\phi_0 , \phi_\infty\}$, $M_\phi := \max\{\phi_0 , \phi_\infty\}$, $m_v := \min\{0 , v_0\}$ and $M_v := \max\{0 , v_0\}$. By Proposition~\ref{prop-bounds} we have that the maximal solution of  \eqref{eq:System}--\eqref{eq:ICs} is confined in the compact domain
$$
\mathcal{K} := [m_\phi, M_\phi] \times [m_v, M_v] \subset \mathcal{D} \, ,
$$
and the defining vector field $F$ of the first-order system \eqref{eq:System} is $C^\infty (\mathcal{D},\mathbb{R}^2)$. From the extension theorem (see \cite[Corollary~2.16]{Teschl2012}) we conclude that $T_\ast$ cannot be finite, i.e. $T_\ast = + \infty$, and the solution is hence unique on $[0,+\infty)$. Indeed, a solution to a sufficiently regular system of differential equations can only stop existing in finite time $T_\ast$ if 
at least one of its components becomes unbounded or the solution comes arbitrarily close to the boundary of its domain (in this case $\partial \mathcal{D} = \{0\} \times \mathbb{R}$). If it stays bounded and safely away from the boundary, it must exist forever. In this specific case we have that, for any fixed initial condition $(\phi_0, v_0)$ we have that $\phi (t) \geq m_\phi > 0$, and hence $(\phi, v)$ remains at a strictly positive distance $m_\phi$ from the boundary $\partial \mathcal{D} = \{0\} \times \mathbb{R}$. This concludes the proof of the global (future) existence and uniqueness.

Let us now observe that $H$, as in Eq.~\eqref{eq:H}, is strictly positive and continuous on the compact domain $\mathcal{K}$, hence by Weierstrass' extreme value theorem we have that it has a minimum on $\mathcal{K}$. We therefore define 
\begin{equation} \label{eq:Hmin}
    h := \min_{\mathcal{K}} H > 0\, .
\end{equation}
This implies that
$$
H \geq h \,\,\, \Rightarrow \,\,\,
\frac{\dot{a}}{a} \geq h \, ,
$$
and hence 
\begin{equation}\label{eq:aexpbound}
a(t) \geq {\rm e}^{ht} \, ,  \quad \forall t \geq 0 \, .
\end{equation}
Furthermore, from Eq.~\eqref{eq:scalar} we can then infer that
\begin{equation} \label{eq:phiexp}
    \begin{split}
    \ddot{\phi} + 3H \, \dot{\phi} = 0 \,\,\, \Rightarrow \,\,\, 
    \dot{\phi} (t) = v_0 \, {\rm e}^{-3 \int_0^t H(\tau) \, {\rm d} \tau} \, ,
    \end{split}
\end{equation}
and hence
\begin{equation}
\label{eq:phiexpbound}
|\dot{\phi}(t)| \leq |v_0| \, {\rm e}^{-3ht} \, ,  \quad \forall t \geq 0 \, .
\end{equation}
Eq.~\eqref{eq:phiexpbound} implies the limit in \eqref{lim-dphi}.

By Lemma~\ref{vneg} we have that $\phi$ is strictly monotonic (or constant if $v_0=0$) and from Proposition~\ref{prop-bounds} it is also bounded for all $t \geq 0$. Hence by the monotone convergence theorem we have that $\exists \phi_\ast \in [m_\phi, M_\phi]$ such that 
$$
\lim_{t \to + \infty} \phi (t) = \phi_\ast \, .
$$
Recalling the first integral \eqref{eq:first-int} and its properties we find that
$$
I(\phi_0, v_0) = I(\phi(t), v(t)) \xrightarrow{t \to +\infty} I(\phi_\ast, 0) \, ,
$$
namely
$$
I(\phi_0, v_0) =I(\phi_\ast, 0) = \frac{3}{2} \log \left( \frac{\phi_\ast}{\phi_0}\right) \, .
$$
Solving for $\phi_\ast$ yields
$$
\phi_\ast = \phi_0 \, {\rm e}^{\frac{2}{3} \, I(\phi_0, v_0)} \equiv \phi_\infty \, ,
$$
thus concluding the proof of the limit in \eqref{lim-phi}.

Lastly, since $\dot{\phi}/\phi = v/\phi \to 0$ as $t \to + \infty$,
the limit in \eqref{lim-H} follows immediately from Eq.~\eqref{eq:H} and \eqref{lim-phi}.
\end{proof}

We can now derive estimates on the distance between the asymptotic value of the solution and and its value at a time $t \in [0,+\infty)$. Such estimates provide benchmarks for checking the consistency and late-time convergence of numerical solutions.
\begin{prop}
Under the assumptions of Theorem~\ref{th-sol}, the following estimates hold for the solution of \eqref{eq:System}--\eqref{eq:ICs}:
\begin{eqnarray}
|\phi(t)-\phi_\infty|
&\leq& \frac{|v_0|}{3h} \, {\rm e}^{-3h t} \, ,\\
|\dot{\phi}(t)| &\leq& |v_0| \, {\rm e}^{-3ht} \, , \\
|H(t) - H_\infty| &=& O\left( {\rm e}^{-3ht} \right) \, ,
\end{eqnarray}
$\forall t \geq 0$ and with $h = \min_{\mathcal{K}} H$.
\end{prop}
\begin{proof}
    The second inequality has already been shown in the proof of Theorem~\ref{th-sol}.
    
    For the first inequity, we observe that from the fundamental theorem of calculus we have that
    $$
    \phi(\infty)-\phi(t)=\int_t^\infty\dot{\phi}(\tau)\,\d \tau \, ,
    $$
    hence, taking advantage of the second inequality, we have that
    $$
    |\phi(t)-\phi_\infty|
    \leq \int_t^\infty |v_0| \, {\rm e}^{-3h\tau}\,\d\tau
    =\frac{|v_0|}{3h} \, {\rm e}^{-3ht} \, .
    $$
    To prove the last inequality let us first observe that $\mathcal{K}$ is a convex set and that $H \in C^1(\mathcal{D})$. Thus, let us consider the line segment
    $$
    \gamma : [0,1] \to \mathcal{K} \, , \quad \gamma(s) := \big(\phi_\infty+s[\phi(t)-\phi_\infty],\,s\,v(t)\big)
    $$
    connecting $\gamma(0) = (\phi_\infty,0)$ and $\gamma(1) =(\phi(t),v(t))$. Since $\mathcal{K}$ is convex, $\gamma(s) \in \mathcal{K}$ for all $s \in [0,1]$. Since $H \in C^1(\mathcal{D})$, the function $H(\gamma (s))$ is continuously differentiable on $[0,1]$ and by the chain rule we have that
    $$
    \frac{\d}{\d s} \big[ H(\gamma (s)) \big] = \partial_\phi H(\gamma(s))\, [\phi(t)-\phi_\infty] + \partial_v H(\gamma(s)) \, v(t) \, .
    $$
    Since again $H \in C^1(\mathcal{D})$, its partial derivatives are continuous on $\mathcal{D}$ and hence on $\mathcal{K} \subset \mathcal{D}$. Then by compactness of $\mathcal{K}$ and by Weierstrass' extreme value theorem we have that $ \partial_\phi H$ and $ \partial_v H$ attain maximum values on $\mathcal{K}$. Let $$
    \bar{M}_\phi:=\max_{\mathcal K}|\partial_\phi H|\, , \quad \bar{M}_v:=\max_{\mathcal K}|\partial_v H| \, ,
    $$
    then 
    \begin{equation*}
    \begin{aligned}
    &|H(\phi(t),v(t))-H(\phi_\infty,0)|
    = \left|\int_0^1 \frac{\d}{\d s} \big[ H(\gamma (s)) \big] \, \d s \right|\\
&\leq \int_0^1
\bigg[
|\partial_\phi H(\gamma(s))|\,|\phi(t)-\phi_\infty|
+|\partial_v H(\gamma(s))|\,|v(t)|
\bigg] \, \d s\\
&\le\int_0^1
\left[\bar{M}_\phi \, |\phi(t)-\phi_\infty|+\bar{M}_v \, |v(t)|\right] \, \d s\\
&=\bar{M}_\phi \, |\phi(t)-\phi_\infty|+\bar{M}_v \, |v(t)| \, .
\end{aligned}
 \end{equation*}
Then, using the first two estimates we find
$$
|H(t)-H_\infty|\leq |v_0|\left(\frac{\bar{M}_\phi}{3h}+\bar{M}_v\right) \, {\rm e}^{-3ht} \, ,
$$
which implies $|H(t)-H_\infty| = O\left( {\rm e}^{-3ht} \right)$.
\end{proof}

Note that these results could also have been inferred from the exact solutions discussed in \cite{Kamenshchik:2013dga}. Those solutions are expressed explicitly in terms of a convenient parametric time, while their dependence on cosmic time is determined by a quadrature followed by inversion.

\begin{figure}
    \centering
    \includegraphics[width=0.8\linewidth]{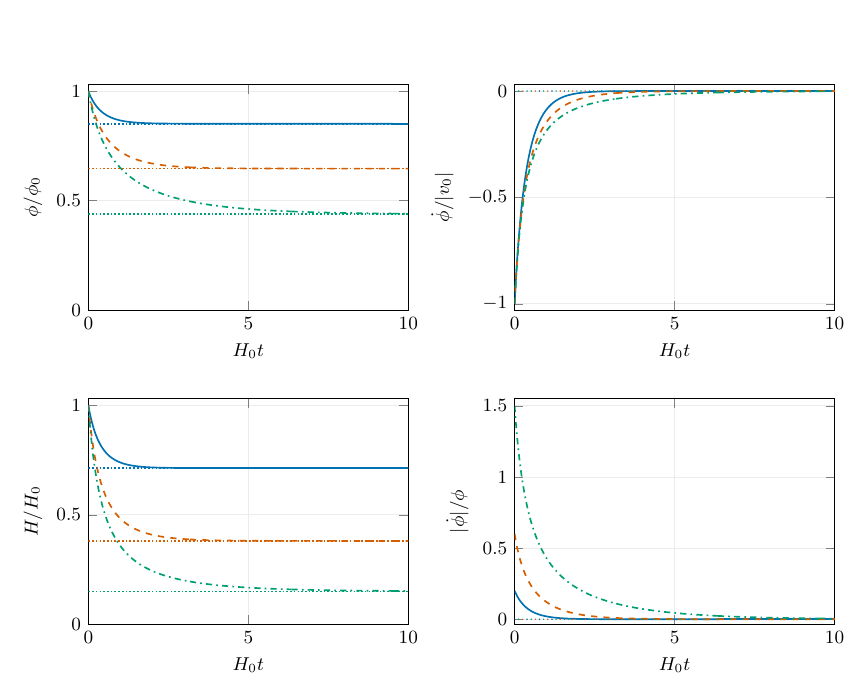}
    \caption{Numerical solutions of the system \eqref{eq:System}--\eqref{eq:ICs} obtained, via the classic 4th-order Runge--Kutta (RK4) method, for $(\phi_0, v_0) = (1,-0.2)$ (solid), $(1,-0.6)$ (dashed), and $(1,-1.5)$ (dot-dashed). Dotted lines indicate the asymptotic limits given by Theorem~\ref{th-sol}.}
    \label{fig:1}
\end{figure}

Figure~\ref{fig:1} illustrates numerical solutions of \eqref{eq:System}--\eqref{eq:ICs}, corroborating the analytical results of this section.

\section{Discussion}
\label{sec:discussion}
Let us restrict our attention to the case of a future-oriented scalar-field gradient; this requires $\dot{\phi} < 0$ and the corresponding BD temperature reads
\begin{equation} \label{eq:T-future}
\K \T = \frac{|\dot{\phi}|}{\phi} > 0 \, .   
\end{equation} 
Based on the analysis in Sec.~\ref{sec:dS} we have to restrict our attention to the system \eqref{eq:System}--\eqref{eq:ICs} with $v_0 < 0$, which guarantees a solution with a scalar field compatible with the requirement of future-orientedness of $\nabla^a \phi$ (see Lemma~\ref{vneg}).

In a FLRW geometry, the expansion scalar reads 
$$\Theta = 3H \, .$$ 
Hence, comparing \eqref{eq:T-future} with \eqref{eq:H} yields
\begin{equation}
H = \frac{\K \T}{2} + \frac{1}{2} \sqrt{\left( 1 + \frac{2}{3} \omega \right) \left(\K \T \right)^2 + \frac{2}{3} \, V_0 \, \phi} \, ,
\end{equation}
which implies (for $\omega, V_0, \phi_0 > 0$ and $v_0 < 0$)
\begin{equation}
\begin{split}
\Theta &= 3H = \frac{3}{2} \, \K \T + \frac{3}{2} \sqrt{\left( 1 + \frac{2}{3} \omega \right) \left(\K \T \right)^2 + \frac{2}{3} \, V_0 \, \phi} \\ 
&>  \,  \chi (\omega) \, \K \T \, > \,\K \T \, ,
\end{split} \label{ineq}
\end{equation}
for all $t \geq 0$, where
\begin{equation}
\chi (\omega) := \frac{3}{2} \left( 1 +  \sqrt{1 + \frac{2}{3} \omega} \right) \, ,
\end{equation}
Both inequalities are strict, since $\phi(t)>0$ by
Theorem~\ref{th-sol} and $(\K\T)(t)>0$ for every finite $t\geq0$. They have a very important consequence for the thermal view of {\em vacuum} induced gravity. Indeed, combining the generalized heat equation \eqref{eq:heat-vacuum} with \eqref{ineq} and Theorem~\ref{th-sol}, we find that 
\begin{equation}
    \lim_{t \to +\infty} (\K \T) (t) = 0 \, .
\end{equation}
Indeed, 
by (\ref{ineq}) Eq.\eqref{eq:heat-vacuum} can be rewritten as
\begin{equation}
\frac{\d(\K\T)}{\d\tau}
=\K\T(\K\T-\Theta) \leq - \big(\chi (\omega) - 1 \big) \, (\K\T)^2 \, ,
\end{equation}
and hence
\begin{equation}
\begin{split}
0 &\leq (\K\T) (t) \\
&\leq \frac{(\K\T) (0)}{1+(\chi (\omega)-1) [(\K\T) (0)] \, t}
\xrightarrow{t \to +\infty} 0 \, .
\end{split}
\end{equation}

Hence, the thermodynamic formalism ``predicts'' that the evolution of the system will make {\em vacuum} induced gravity converge toward GR (plus a cosmological constant) at late times. Thus, the model has an attractor-to-GR mechanism embedded in its dynamics for an expanding, spatially flat FLRW universe. This is consistent with Theorem~\ref{th-sol}, which establishes convergence toward this configuration. We have therefore provided two complementary ways of inferring the attractor-to-GR mechanism for this model: the thermodynamics of scalar-tensor gravity (corroborating the analysis of \cite{Giusti:2026ymb}) and a rigorous analysis of the evolution equations of the system.

\begin{figure}
    \centering
    \includegraphics[width=0.4\linewidth]{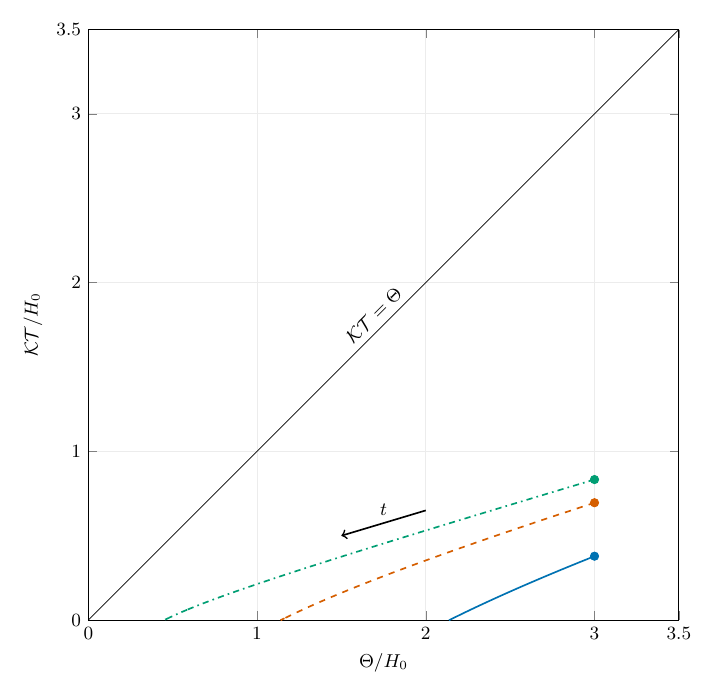}
    \caption{Trajectories of \eqref{eq:System}--\eqref{eq:ICs} in the $(\Theta,\K\T)$--plane computed via classical RK4 for $(\phi_0, v_0) = (1,-0.2)$ (solid), $(1,-0.6)$ (dashed), and $(1,-1.5)$ (dot-dashed). The arrow indicates increasing cosmic time $t$.}
    \label{fig:2}
\end{figure}

It is also useful to visualize this analysis in the $(\Theta,\K\T)$--plane (see Figure~\ref{fig:2}).

\section{Conclusions}
\label{sec:conclusions}
We have shown that every solution on the expanding branch of {\em vacuum} induced gravity, with a positive initial scalar field, exists for all future cosmic times and converges to a de Sitter state. The proof relies on a first integral that confines the scalar field between strictly positive bounds, thus excluding both finite-time blow-ups and an asymptotic approach to the singular boundary. This procedure also determines the late-time constant asymptotic values of the scalar field, its time derivative, and of the Hubble parameter. The asymptotic behaviour established here is consistent with earlier investigations and can also be recovered from known exact parametric solutions.

Restricting the analysis to a future-directed scalar-field gradient, we find that the temperature of the considered model is positive and the Friedmann constraint equation confines the evolution to the region $\K \T < \Theta$ at all times during the evolution of the system. The generalized heat equation therefore describes a ``monotonic cooling of induced gravity'' toward the GR equilibrium state, while the global analysis implies that $\K\T$ tends to zero as $t \to + \infty$. The late-time de Sitter state thus corresponds to the zero-temperature GR equilibrium, corroborating the thermodynamic analysis of induced gravity developed in \cite{Giusti:2026ymb}.

\section*{Acknowledgments}
\noindent We are grateful to A. Tronconi, A. Kamenchtchik, R. Casadio, and M. Sebastianutti for valuable discussions. 

\noindent A.G.~is supported by the Italian Ministry of Universities and Research (MUR) through the grant ``BACHQ: Black Holes and The Quantum'' (grant no. J33C24003220006) and by the INFN grant FLAG. G.F.~is supported, in part, by the INFN grant GeoSymQFT.

\noindent We also acknowledge the COST Action CA23115 (RQI).

\noindent This work has been carried out in the framework of activities of the National Group of Mathematical Physics (GNFM, INdAM).

\onecolumngrid

\twocolumngrid

%
%
%
\bibliographystyle{apsrev4-2}
\bibliography{Biblio}
\end{document}